\documentclass[aps,prl,preprint,groupedaddress,showkeys]{revtex4-1}

\usepackage{graphicx}
\usepackage{amsmath}
\usepackage{bm}

\begin{document}


\title{An all-dielectric bowtie waveguide with deep subwavelength mode confinement}


\author{Wencheng Yue, Peijun Yao, and Lixin Xu}
 \affiliation{Department of Optics and Optical Engineering, University of Science and Technology of China, Hefei 230026, China}


\date{\today}

\begin{abstract}
To fulfil both size and power requirements for future photonic integrated circuits, an effective approach is to miniaturize photonic components. Surface plasmon polariton (SPP) is one of the most promising candidates for subwavelength mode confinement, however, structures based on SPP are subject to inevitable high propagation loss. Here, we report an all-dielectric bowtie (ADB) waveguide consisting of two identical silicon wedges embedded in a silica cladding with a nanoscale gap. Because of successive slot and antislot effects, the gap behaves as a `capacitor-like' energy storage that makes the ADB waveguide have similar or even smaller mode area than the hybrid plasmonic waveguides recently reported. What is more important is that the ADB waveguide supports a quasi-TM eigenmode, which is lossless fundamentally because of no metal constituent. This makes our ADB waveguide have essential development in propagation length compared with the plasmonic waveguide. The ADB waveguide is fully compatible with semiconductor fabrication techniques and could give rise to truly nanoscale semiconductor-based photonics.
\end{abstract}

\keywords{Dielectric waveguide, deep subwavelength confinement, nanophotonics, photonic integrated circuits.}

\maketitle

\section{Introduction}
It is of great importance manipulating light at subwavelength scale in nanophotonics \cite{kirchain2007roadmap}. Surface plasmon polariton (SPP) \cite{boardman1982electromagnetic,barnes2003surface} waveguides are among the most promising candidates for their unique capabilities of breaking the diffraction limit and providing tight light confinement in deep subwavelength scale \cite{barnes2003surface,wang2014review,gramotnev2010plasmonics,takahara2004nano,maier2003local}. Recently many hybrid SPP waveguides have been reported, such as hybrid plasmonic waveguides \cite{oulton2008hybrid,dai2009silicon,avrutsky2010sub,bian2010dielectric,zhao2010coaxial}, long-range hybrid plasmonic waveguides \cite{bian2009symmetric,chen2012novel,xiang2013long}, hybrid wedge plasmonic waveguides \cite{bian2011hybrid,bian2014bow}, and long-range hybrid wedge plasmonic waveguides \cite{zhang2014long,ma2014hybrid,ma2015hybrid}. These hybrid SPP waveguides could offer strong mode confinement but suffer inevitable high propagation loss due to the metallic ohmic loss \cite{zia2004geometries}. Conventional dielectric waveguides, such as photonic crystals \cite{cregan1999single,wiederhecker2007field,altug2006ultrafast}, guide light with low loss but are subject to the diffraction limit in each direction. Although subwavelength mode confinement has been found in all-dielectric coupled silicon waveguides \cite{almeida2004guiding,xu2004experimental,almeida2003light,mullner2006structural}, it is only limited to one dimension.

In this letter, we report an all-dielectric bowtie (ADB) waveguide capable of deep subwavelength confinement in two dimensions. The ADB waveguide consists of two identical high-permittivity silicon (Si) wedges embedded in a low-permittivity silica (SiO$_2$) cladding with a nanoscale separation gap. It can confine light inside the nanoscale gap region. This makes our ADB waveguide have similar or even smaller mode area than the hybrid plasmonic waveguides aforementioned. More importantly, the guided mode is an eigenmode of our proposed ADB waveguide \cite{almeida2004guiding,xu2004experimental,almeida2003light,mullner2006structural}, which is fundamentally lossless because of no metal constituent. This is an improvement in nature, compared to the plasmonic waveguide. The advantages of our ADB waveguide-deep subwavelength mode confinement in two dimensions and lossless propagation-make it have extensive applications, such as photonic integrated circuits, high-resolution spatial light modulators, nonlinear optics, and so on.
\section{Theoretical background}
The mode area ($A_{eff}$) of an optical waveguide is given by the ratio of the total electromagnetic energy to the maximum electromagnetic energy density \cite{oulton2008hybrid},
\begin{equation}
\label{eq.1}
A_{eff}=\frac{1}{max\{W(\bm{r})\}}\int\!\!\!\!\int{W(\bm{r})d^2\bm{r}},
\end{equation}
where $W(\bm{r})$ is the electromagnetic energy density. For non-dispersion material
\begin{equation}
\label{eq.2}
W(\bm{r})=\frac{1}{2}[\epsilon(\bm{r})|\bm{E}(\bm{r})|^2+\mu_0|\bm{H}(\bm{r})|^2].
\end{equation}
Obviously, the mode area is determined by the electromagnetic energy density at the position where it is maximized and can be greatly shrunk by increasing the maximum electromagnetic energy density term in the denominator of Eq. (1). The all-dielectric slot waveguides ever reported \cite{almeida2004guiding,xu2004experimental,almeida2003light,mullner2006structural} could realize subwavelength mode confinement in the direction perpendicular to the slot interfaces by squeezing light in the low-permittivity slot region. The light confinement originates from the electromagnetic boundary condition on the normal component of the electric field ($E_n$).  Electromagnetic boundary conditions suggest that the normal component of the electric displacement field ($D_n$) and the tangential component of the magnetic field ($H_t$) are continuous across an interface between two media, i.e., $D_{n, high}$ = $D_{n, low}$, $H_{t, high}$ = $H_{t, low}$. $D_{n, high}$ and $D_{n, low}$ ($H_{t, high}$ and $H_{t, low}$) are the normal components of the electric displacement field (the tangential components of the magnetic field) in the high and low permittivity materials, respectively. The corresponding normal component of the electric field ($E_n$ = $D_n$/$\epsilon$) is discontinuous across the interface, $E_{n,low}$ = $D_{n,low}$/$\epsilon_{low}$ = ($\epsilon_{high}$/$\epsilon_{low}$)$E_{n,high}$, where $\epsilon_{high}$ and $\epsilon_{low}$ are respectively the permittivities of the high and low permittivity media. For a slot waveguide as shown in Fig. 1(a) (Top), the electric and magnetic fields are dominated by their normal and tangential components respectively \cite{almeida2004guiding,xu2004experimental,almeida2003light,mullner2006structural}. So the electromagnetic energy densities near the interface between the low and high permittivity materials are respectively
\begin{equation}
\label{eq.3}
W_{low}=\frac{1}{2}[\epsilon_{low}|E_{n, low}|^2+\mu_0|H_{t, low}|^2]=\frac{1}{2}[\frac{\epsilon_{high}^2}{\epsilon_{low}}|E_{n, high}|^2+\mu_0|H_{t, high}|^2],
\end{equation}
\begin{equation}
\label{eq.4}
W_{high}=\frac{1}{2}[\epsilon_{high}|E_{n, high}|^2+\mu_0|H_{t, high}|^2],
\end{equation}
where $W_{low}$ and $W_{high}$ correspond to the electromagnetic energy densities in the low and high permittivity materials respectively. Obviously, the electric energy density in the slot is larger than that in the high-permittivity material. In view of the fact that the electric energy density is much larger than the magnetic energy density in the slot, the slot will squeeze light into nanoscale low-permittivity material surrounded by high-permittivity material [Fig. 1(a) Bottom], which is also known as slot effect \cite{almeida2004guiding}.

\begin{figure}[!t]
\centering
\includegraphics[width=0.85\textwidth]{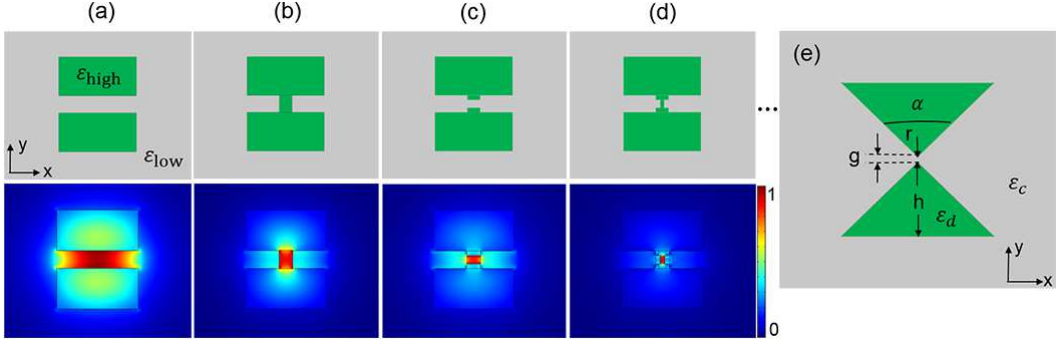}
\caption{(a)-(d) Structures (Top) and normalized electromagnetic energy density distributions (Bottom) of different slot waveguides, where the antislots and slots are introduced in turn. The width and height of the green rectangular waveguide are 300 nm and 150 nm, respectively. The width of the slot or antislot is 65 nm, 50 nm, 30 nm, and 15 nm in sequence from (a) to (d). (e) Schematic illustration of our proposed ADB waveguide. The center of the ADB waveguide defines the origin.}
\label{Fig1}
\end{figure}

Actually, the mode area could be further reduced by taking advantaging of the second electromagnetic boundary condition on the tangential component of the electric field ($E_t$) \cite{choi2017self,hu2016design}. The tangential component of the electric field ($E_t$) and the normal component of the magnetic flux density ($B_n$) are continuous across an interface between two media, i.e., $E_{t, high}$ = $E_{t, low}$, $B_{n, high}$ = $B_{n, low}$, where $E_{t, high}$ and $E_{t, low}$ ($B_{n, high}$ and $B_{n, low}$) are the tangential components of the electric field (the normal components of the magnetic flux density) in the high and low permittivity materials, respectively. Due to $\mu$ = $\mu_0$  for nonmagnetic materials, the corresponding normal component of the magnetic field ($H_n$ = $B_n$/$\mu$ = $B_n$/$\mu_0$) is also continuous across the interface, i.e., $H_{n,high}$ = $H_{n,low}$. Considering the waveguide structure as shown in Fig. 1(b) (Top), which is constituted by introducing a high-permittivity antislot in the slot waveguide, the electromagnetic energy densities near the interface between the high-permittivity antislot and the low-permittivity slot are respectively
\begin{equation}
\label{eq.5}
W_{antislot}=\frac{1}{2}[\epsilon_{high}|E_{t, high}|^2+\mu_0|H_{n, high}|^2]=\frac{1}{2}[\frac{\epsilon_{high}}{\epsilon_{low}}\epsilon_{low}|E_{t, low}|^2+\mu_0|H_{n, low}|^2],
\end{equation}
\begin{equation}
\label{eq.6}
W_{low}=\frac{1}{2}[\epsilon_{low}|E_{t, low}|^2+\mu_0|H_{n, low}|^2],
\end{equation}
where $W_{antislot}$ and $W_{low}$ correspond to the electromagnetic energy densities in the antislot and slot respectively. Likewise, the antislot confines light into nanoscale high-permittivity material [Fig. 1(b) Bottom], named as antislot effect \cite{hu2016design}.

The mode can be ceaselessly squeezed by alternatively introducing a series of high-permittivity antislot and low-permittivity slot into the slot waveguide, as depicted in Fig. 1. The successive slot and antislot effects, arising from the orthogonal nature of the electromagnetic boundary conditions, can progressively and respectively squeeze light in the in-plane perpendicular direction ($y$-direction) and the horizontal direction ($x$-direction), and light is mainly confined to the last antislot or slot, as shown in Fig. 1(Bottom). Figure 1 shows how we come up with the idea.

\section{Waveguide structure and mode properties}
The designed ADB waveguide, illustrated in Fig. 1(e), consists of two identical high-permittivity Si wedges embedded in a low-permittivity SiO$_2$ cladding, the separation between the two wedges is indicated by $g$. In the following study, we change the wedge height ($h$), wedge tip angle ($\alpha$), and the gap distance ($g$) between the two Si wedges to control the effective mode area and electromagnetic field distribution of the quasi-TM eigenmode of our ADB waveguide at the telecommunications wavelength ($\lambda$ = 1,550 nm). Figure 2(a) shows the dependence of normalized mode area ($A_{eff}$/$A_0$) on $h$ and $\alpha$ at $g$ = 2 nm, where $A_0$ is the diffraction-limited mode area and defined as $\lambda^2$/4, $A_{eff}$ is calculated by the finite element method (FEM)\cite{reddy1993introduction}. For a small wedge height and tip angle ($h <$ 180 nm, $\alpha$ $<$ 80$^{\circ}$), the electromagnetic energy of the quasi-TM eigenmode of our ADB waveguide diffuses in the cladding [Fig. 2(b)], which results in a large mode area [Fig. 2(a)]. Inversely, a large wedge height ($h >$ 300 nm) leads to the electromagnetic energy spreading to the Si wedges [Figs. 2(c) and 2(d)], accompanied by a large mode area [Fig. 2(a)]. At moderate wedge height (180 nm $< h <$ 300 nm), the electromagnetic energy of the quasi-TM mode is strongly confined within the gap region in two dimensions [Figs. 2(e)-2(g)], accounting for the minimum mode areas [Fig. 2(a)]. The white arrows in Figs. 2(e)-2(g) indicate the in-plane electric vector distributions of the guided mode of our ADB waveguide. Clearly, the major electric field component is parallel to $y$ direction, indicating the guided mode of the ADB waveguide is quasi-TM mode. Furthermore, the quasi-TM mode is an eigenmode of the proposed ADB waveguide \cite{almeida2004guiding,xu2004experimental,almeida2003light,mullner2006structural}, and there is theoretically no loss because of no metal constituent, which is consistent with the negligible imaginary part of the effective refractive index calculated by FEM. The lossless propagation makes the ADB waveguide have naturally superiority to the plasmonic waveguide. What is more interesting here is that, despite this lossless propagation, the ADB waveguide provides strong subwavelength mode confinement [Fig. 2(a)].

\begin{figure}[!t]
\centering
\includegraphics[width=0.29\textwidth]{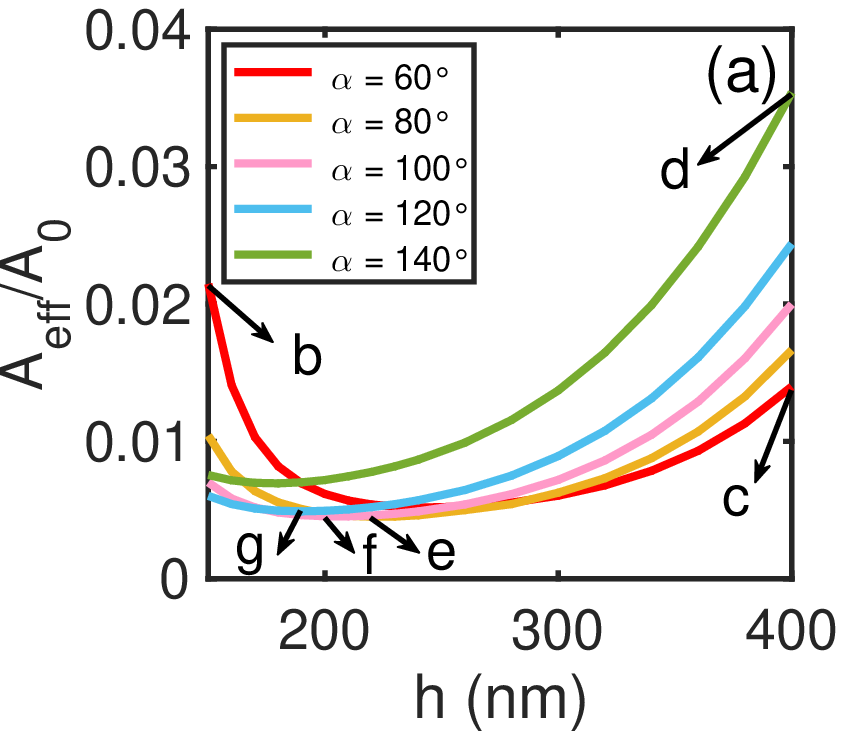}
\label{Fig2a}
\includegraphics[width=0.5\textwidth]{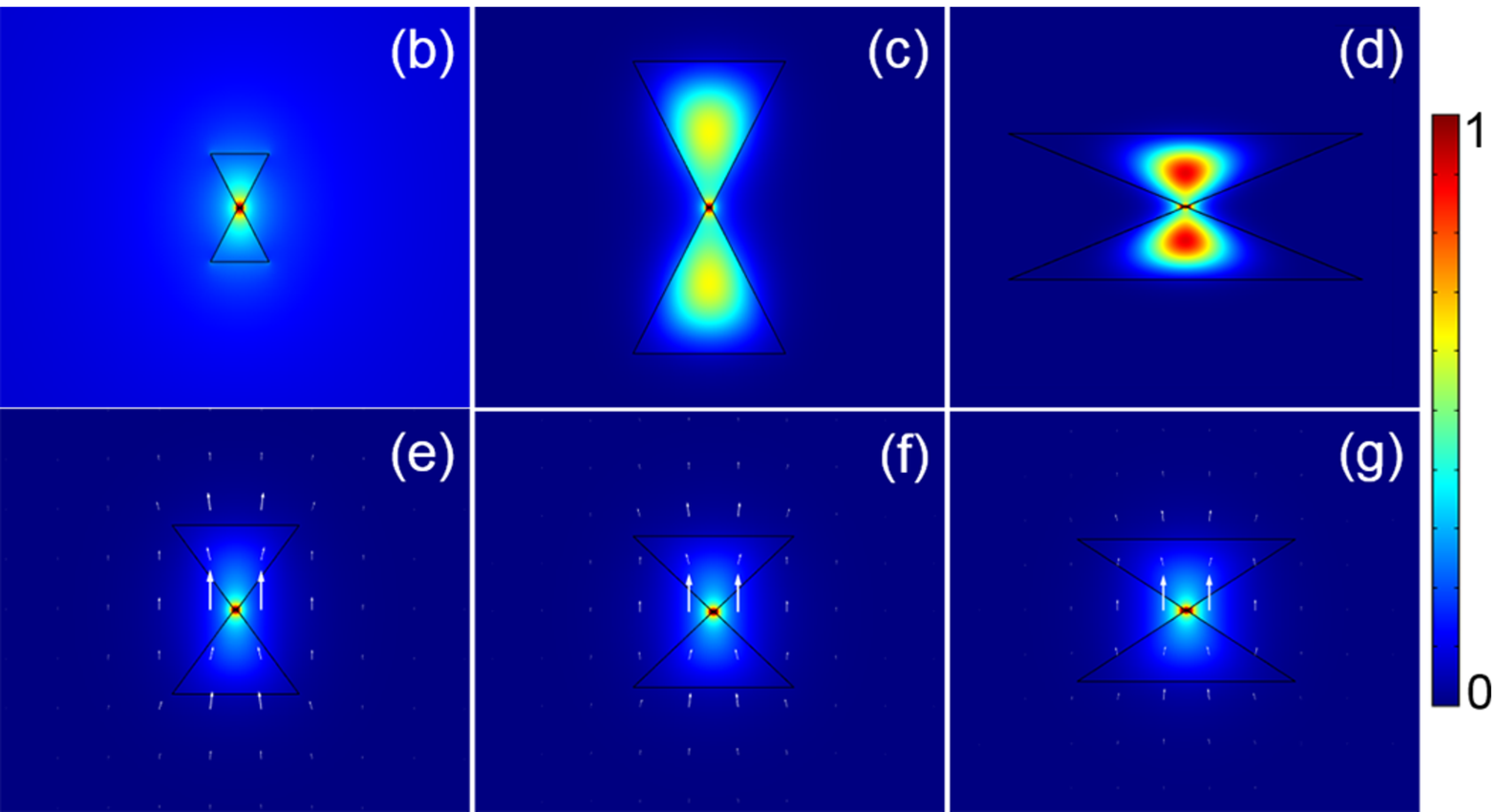}
\label{Fig2b-g}
\caption{(a) Normalized mode area ($A_{eff}$/$A_0$) versus wedge height ($h$) at different wedge tip angle ($\alpha$) for $g$ = 2 nm. (b)-(g) Normalized electromagnetic energy density distributions for [$h$, $\alpha$] = [150 nm, 60$^{\circ}$], [$h$, $\alpha$] = [400 nm, 60$^{\circ}$], [$h$, $\alpha$] = [400 nm, 140$^{\circ}$], [$h$, $\alpha$] = [220 nm, 80$^{\circ}$], [$h$, $\alpha$] = [200 nm, 100$^{\circ}$], and [$h$, $\alpha$] = [190 nm, 120$^{\circ}$], corresponding to the points indicated in panel (a). The white arrows in (e)-(g) indicate the in-plane electric vector distributions.}
\end{figure}

Figure 3(a) depicts the dependence of the normalized mode area ($A_{eff}$/$A_0$) on gap ($g$) for the same set of the wedge height ($h$) and wedge tip angle ($\alpha$) as Figs. 2(e)-2(g). The normalized mode area increases along with the gap and it achieves minimum values at $g$ = 0 nm which corresponds to the case that the ADB waveguide ends with antislot. The gap provides the means to store electromagnetic energy, giving rise to subwavelength mode confinement in two dimensions. This is further certified by the distribution of electromagnetic energy density at $x$ = 0 or $y$ = 0. Evidently, the electromagnetic energy density decreases sharply in two Si wedge regions, as shown in Figs. 3(b)-3(e), demonstrating a subwavelength mode confinement in $y$ direction for all the considered $g$. The inset in Fig. 3(f) shows the full-width at half-maximum (FWHM) of the mode at $y$ = 0, indicating a subwavelength mode confinement in $x$ direction. In all, the designed ADB waveguide can provide subwavelength mode confinement in both in-plane directions.

\begin{figure}[!t]
\centering
\includegraphics[width=0.29\textwidth]{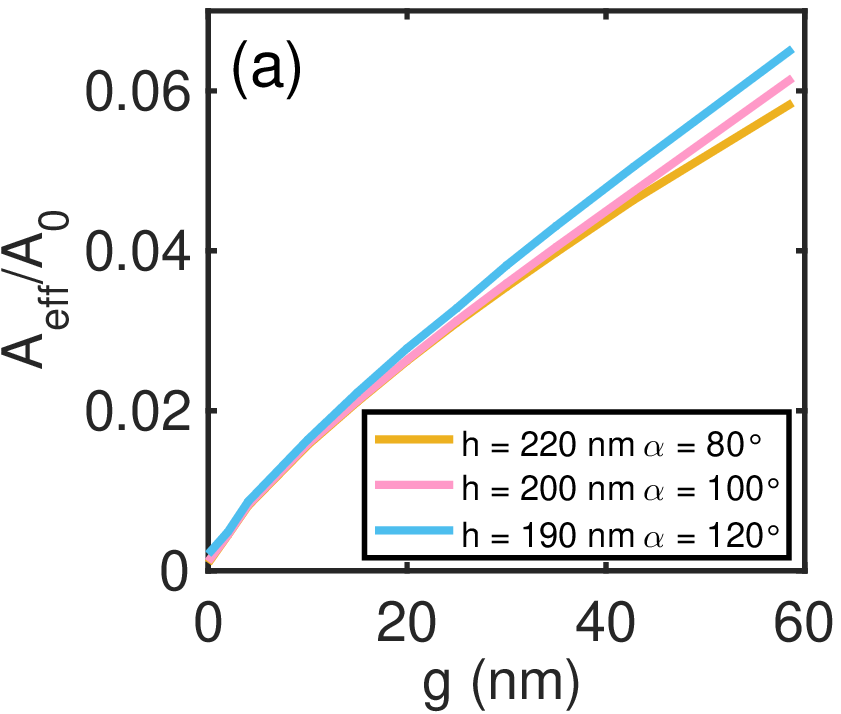}
\label{Fig3a}
\includegraphics[width=0.29\textwidth]{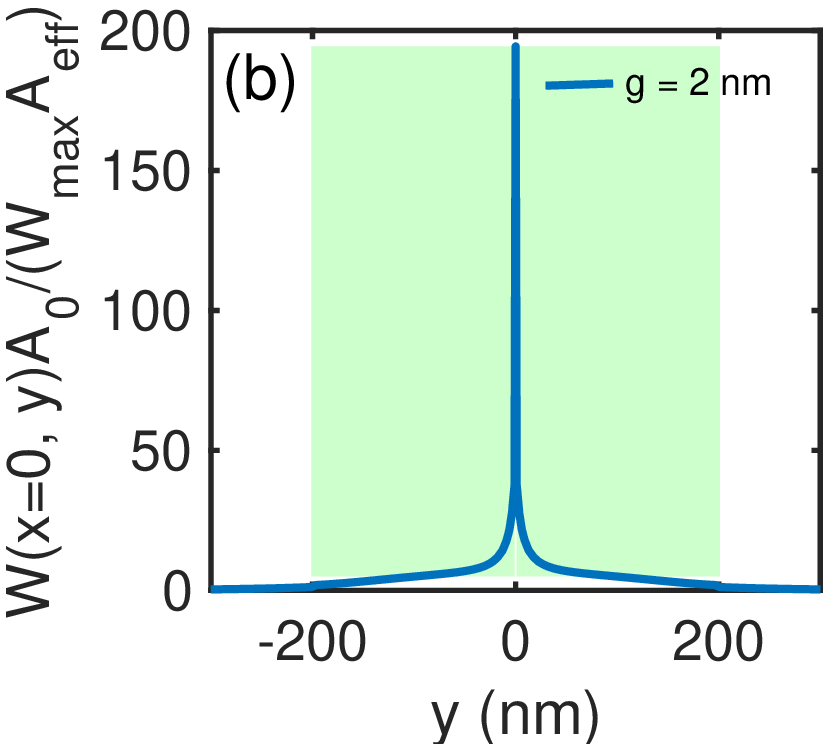}
\label{Fig3b}
\includegraphics[width=0.29\textwidth]{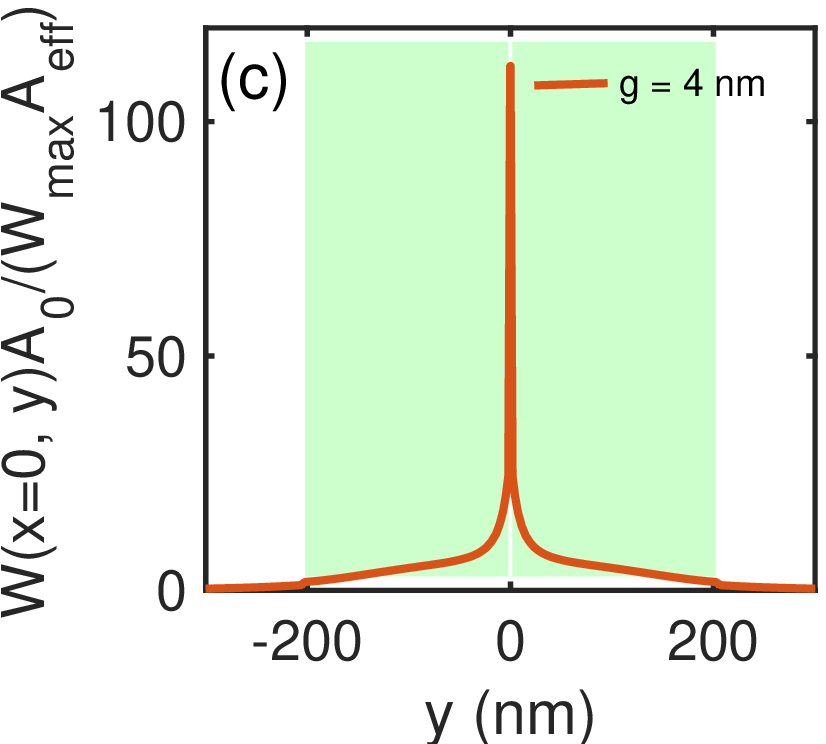}
\label{Fig3c}

\includegraphics[width=0.29\textwidth]{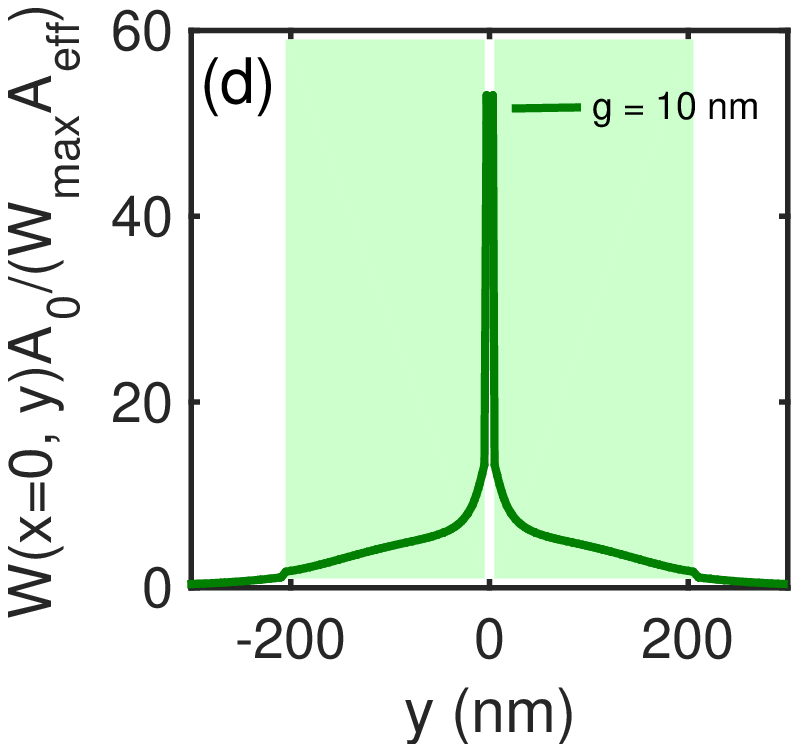}
\label{Fig3d}
\includegraphics[width=0.29\textwidth]{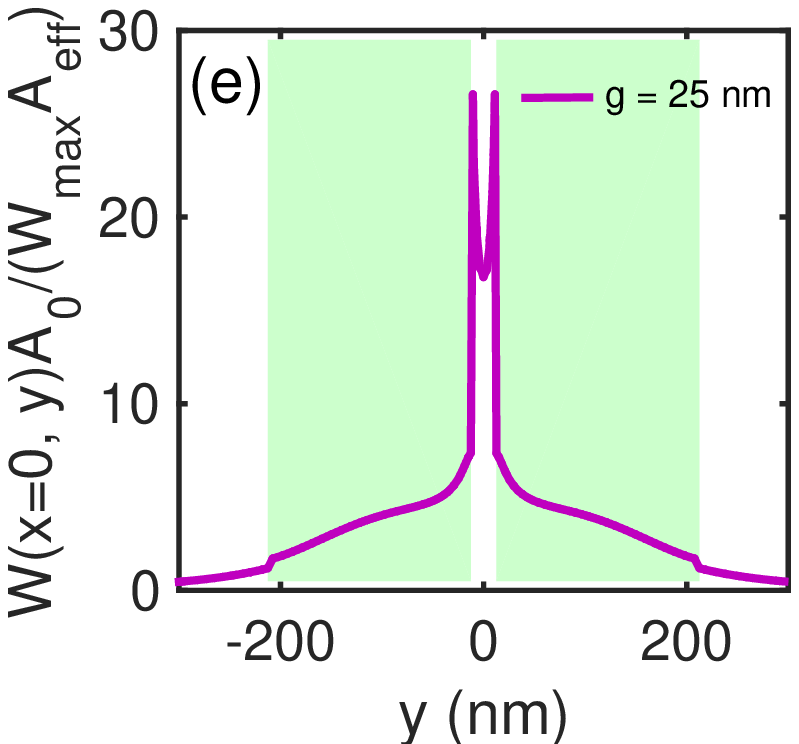}
\label{Fig3e}
\includegraphics[width=0.29\textwidth]{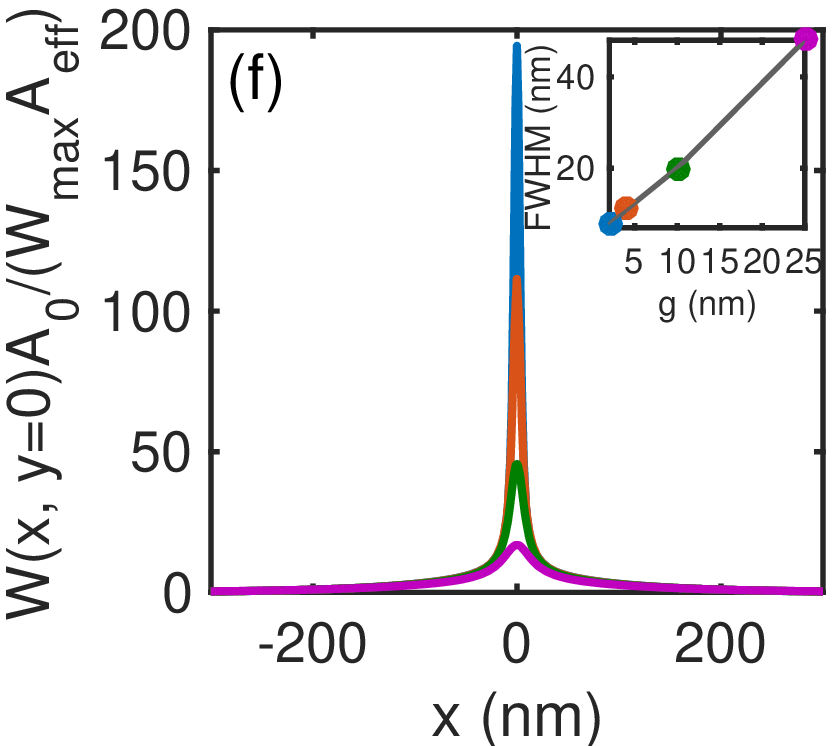}
\label{Fig3f}
\caption{(a) Normalized mode area ($A_{eff}$/$A_0$) versus gap ($g$) for the same set of $h$ and $\alpha$ as Figs. 2(e)-2(g). (b)-(e) Electromagnetic energy density at $x$ = 0 for $g$ = 2 nm, $g$ = 4 nm, $g$ = 10 nm, and $g$ = 25 nm, shows the confinement in the gap region (no shading). The green shaded areas represent two Si wedge regions. (f) Electromagnetic energy density at $y$ = 0 shows subwavelength confinement along $x$ direction. The inset shows the FWHM of the mode at $y$ = 0. The parameters used in (b)-(f) are $h$ = 200 nm, $\alpha$ = 100$^{\circ}$.}
\end{figure}

\begin{figure}[!t]
\centering
\includegraphics[width=0.3\textwidth]{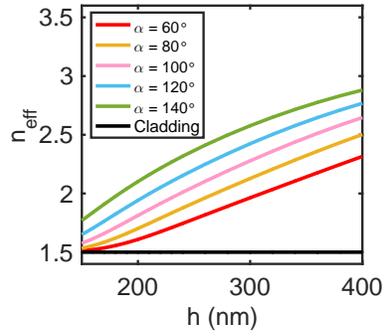}
\caption{Effective refractive index ($n_{eff}$) of the quasi-TM eigenmode versus wedge height ($h$) for different tip angle ($\alpha$) when $g$ = 2 nm. The black line represents the cladding refractive index.}
\label{Fig4}
\end{figure}

In order to get a deeper understanding, we analyzed the dependence of effective refractive index ($n_{eff}$) of the quasi-TM eigenmode on $h$ and $\alpha$ at $g$ = 2 nm (Fig. 4). The parameters used in Fig. 4 is the same as Fig. 2(a). The effective refractive index increases along with the wedge height ($h$) or tip angle ($\alpha$). Smaller wedge height and tip angle make the effective refractive index of the quasi-TM mode closer to the cladding refractive index, resulting in diffused electromagnetic energy in the cladding [Fig. 2(b)]. At certain wedge tip angle, there is always a wedge height where the mode area is minimum [Fig. 2(a)]. Furthermore, the smaller the tip angle, the larger the corresponding wedge height where the minimum mode area is achieved. Table I shows the minimum normalized mode areas and the corresponding wedge height for different wedge tip angle when $g$ = 2 nm. These minimum normalized mode areas are mainly at the order of magnitude of 10$^{-3}$ and keep on the same level with the hybrid plasmonic waveguide at the same gap \cite{oulton2008hybrid}. When $h$ = 200 nm $\alpha$ = 100$^{\circ}$ or $h$ = 220 nm $\alpha$ = 80$^{\circ}$, the smallest normalized mode area of about 4.5$\times$10$^{-3}$ is achievable for $g$ = 2 nm. This mode area is less than the hybrid plasmonic waveguide at the same gap (about 77$\%$) \cite{oulton2008hybrid}. As the gap distance decreases further (g $\longrightarrow$ 0), we can expect that a smaller mode area is available [Fig. 3(a)].

\begin{table}[htbp]
\centering
\caption{\label{table1} Minimum normalized mode areas ($A_{eff}$/$A_0$) and the corresponding wedge height ($h$) for different tip angle ($\alpha$) when $g$ = 2 nm.}
\begin{tabular}{c|c|c|c|c|c|c|c|c}
\hline
\hline
$\alpha$ & 20$^{\circ}$ & 40$^{\circ}$ & 60$^{\circ}$ & 80$^{\circ}$ & 100$^{\circ}$ & 120$^{\circ}$ & 140$^{\circ}$ & 160$^{\circ}$\\
\hline
$h$ (nm) & 330 & 280 & 240 & 220 & 200 & 190 & 180 & 170\\
\hline
$A_{eff}$/$A_0$ & 11.2$\times$10$^{-3}$ & 7.3$\times$10$^{-3}$ & 5.2$\times$10$^{-3}$ & 4.5$\times$10$^{-3}$ & 4.5$\times$10$^{-3}$ & 4.9$\times$10$^{-3}$ & 6.9$\times$10$^{-3}$ & 12.3$\times$10$^{-3}$\\
\hline
\hline
\end{tabular}
\end{table}

\section{Conclusions}
Our work shows that it is possible for all-dielectric waveguides to confine light in a deep subwavelength scale along with a fundamentally lossless propagation. The designed ADB waveguide, corresponding to introducing an infinite number of interlocked antislot and slot in the slot waveguide, can realize deep subwavelength mode confinement in two dimensions and the smallest normalized mode area of 4.5$\times$10$^{-3}$ is achieved when $g$ = 2 nm. Moreover, the mode area can be further reduced by decreasing the gap. What is more important is that the supported quasi-TM eigenmode of the ADB waveguide is fundamentally lossless because of no metal constituent. All of these will make our ADB waveguide have extensive application prospects in many fields, such as increasing the integration of optical integrated circuit, improving the efficiency of near-field optical probes, enhancing the fluorescence signals in bioimaging, and increasing the sensibility of optical sensing, and so on.

\begin{acknowledgments}
This work was supported by National Key Basic Research Program of China (2012CB922003); National Natural Science Foundation of China (NSFC) (61177053); and Anhui Provincial Natural Science Foundation (1508085SMA205, 1408085MKL0).
\end{acknowledgments}


\end{document}